\documentclass[conference]{IEEEtran}

\ifCLASSINFOpdf
\else
\fi
\usepackage{graphicx}
\usepackage{tabularx,booktabs}
\usepackage{dingbat}
\usepackage{diagbox}
\usepackage{multirow} 
\usepackage[hyphens]{url}
\usepackage[hidelinks,breaklinks]{hyperref}
\usepackage{xcolor}
\usepackage{amsfonts, amsmath, amsthm, amssymb}
\usepackage{subcaption}
\hypersetup{breaklinks=true}
\usepackage{tikz}
\usepackage{textcomp}
\usepackage{lipsum}
\usepackage{setspace}

\IEEEoverridecommandlockouts
\newcommand\copyrighttext{%
  \footnotesize \textcopyright 2024 IEEE.  Personal use of this material is permitted.  Permission from IEEE must be obtained for all other uses, in any current or future media, including reprinting/republishing this material for advertising or promotional purposes, creating new collective works, for resale or redistribution to servers or lists, or reuse of any copyrighted component of this work in other works.}
\newcommand\copyrightnotice{%
\begin{tikzpicture}[remember picture,overlay]
\node[anchor=south,yshift=10pt] at (current page.south) {\fbox{\parbox{\dimexpr\textwidth-\fboxsep-\fboxrule\relax}{\copyrighttext}}};
\end{tikzpicture}%
}

\usepackage{graphicx}
\graphicspath{{figures/}}
\begin{document}

\bstctlcite{IEEEexample:BSTcontrol}
%
\title{Performance Evaluation of Uplink 256QAM on Commercial 5G New Radio (NR) Networks}

\author{
    \IEEEauthorblockN{Kasidis Arunruangsirilert\IEEEauthorrefmark{1}, Pasapong Wongprasert\IEEEauthorrefmark{2}, Jiro Katto\IEEEauthorrefmark{1}}
    \IEEEauthorblockA{\IEEEauthorrefmark{1}Department of Computer Science and Communications Engineering, Waseda University, Tokyo, Japan}
    \IEEEauthorblockA{\IEEEauthorrefmark{2}Department of Electrical Engineering, Chulalongkorn University, Bangkok, Thailand
    \\\{kasidis, katto\}@katto.comm.waseda.ac.jp, 6670331021@student.chula.ac.th}
}
\vspace{-2mm}
%
\maketitle
\vspace{-2mm}
\copyrightnotice
\begin{abstract}

 \setstretch{0.9}
While Uplink 256QAM (UL-256QAM) has been introduced since 2016 as a part of 3GPP Release 14, the adoption was quite poor as many Radio Access Network (RAN) and User Equipment (UE) vendors didn't support this feature. With the introduction of 5G, the support of UL-256QAM has been greatly improved due to a big re-haul of RAN by Mobile Network Operators (MNOs). However, many RAN manufacturers charge MNOs for licenses to enable UL-256QAM per cell basis. This led to some MNOs hesitating to enable the feature on some of their gNodeB or cells to save cost.

 \setstretch{0.9}
Since it's known that 256QAM modulation requires a very good channel condition to operate, but UE has a very limited transmission power budget. In this paper, 256QAM utilization, throughput and latency impact from enabling UL-256QAM will be evaluated on commercial 5G Standalone (SA) networks in two countries: Japan and Thailand on various frequency bands, mobility characteristics, and deployment schemes. By modifying the modem firmware, UL-256QAM can be turned off and compared to the conventional UL-64QAM. The results show that UL-256QAM utilization was less than 20\% when deployed on a passive antenna network resulting in an average of 8.22\% improvement in throughput. However, with Massive MIMO deployment, more than 50\% utilization was possible on commercial networks. Furthermore, despite a small uplink throughput gain, enabling UL-256QAM can lower the latency when the link is fully loaded with an average improvement of 7.97 ms in TCP latency observed across various test cases with two TCP congestion control algorithms.
\end{abstract}



%
\IEEEpeerreviewmaketitle

\vspace{-2mm}
\section{Introduction}
 \setstretch{0.941}

After Downlink 256QAM (DL-256QAM) was introduced as a part of 3GPP Release 12 during the 4G Long-Term Evolution (LTE) era \cite{7470807}, there are need for high throughput uplink on the mobile. Uplink carrier aggregation (UL-CA), which was introduced back in the 3GPP Release 10, can increase the maximum achievable uplink throughput by allowing the User Equipment (UE) to transmit the user data via more than one carrier simultaneously. Only a minority of the UE actually implemented such functionality with UE supporting interband uplink carrier aggregation being even more rare. Furthermore, transmitting on two carriers resulted in higher power usage, reducing the battery life of the UE significantly. Therefore, Uplink 256QAM (UL-256QAM) was introduced as a part of 3GPP Release 14 \cite{3GPP_36-306}, promising higher spectral efficiency than the maximum of 64QAM modulation in the earlier releases.

\begin{figure}[!t]
    \centering
    \includegraphics[width=0.98\linewidth]{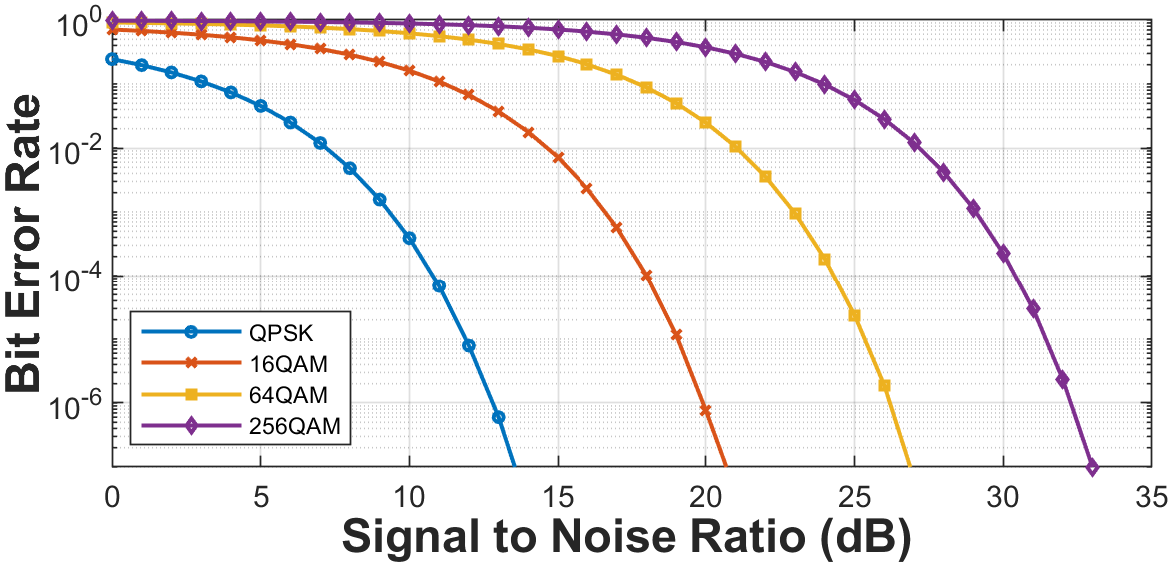}
    \caption{\textbf{Simulation Results:} Bit Error Rate (BER) vs \\ Signal to Noise Ratio (SNR) per each Modulation Scheme}
    \label{fig:ModulationSimulation}
    \vspace{-8mm}
\end{figure}

Theoretically, by utilizing 256QAM modulation, a throughput uplift of 25\% can be achieved over 64QAM over the same channel bandwidth, which can be useful for terrestrial transmission of Ultra-High-Definition (UHD) 4K/8K and 3D point cloud contents \cite{7329950}\cite{tsuchida_2019}, where maximum spectral efficiency is desired. The simulation shows that 256QAM is less resilient to noise and interference, requiring SNR of more than 30 dB to be fully utilized without channel coding and error correction (see Fig. \ref{fig:ModulationSimulation}). While this is not an issue for a unidirectional broadcasting network as all base stations send out the same information simultaneously, causing no interference to each other, in mobile networks, each base station and UE transmit different user data, causing interference to neighbor transmission. Therefore, such a level of SINR is extremely rare in the live commercial RAN network, especially in areas with high cell density. This raises a big concern as DL-256QAM is quite difficult to achieve in the real world despite the base station possessing a high transmission power budget \cite{7129725}. With the UE transmission power limited to 23 dBm \cite{3GPP_36-101}, to comply with Specific Absorption Rate (SAR) requirements, the throughput gain from implementing UL-256QAM is questionable as UE is unlikely to have sufficient power headroom to overcome the interference and achieve high SINR required for 256QAM modulation, and will likely opt for lower modulation index, which is already available in the earlier LTE releases. This led to poor adoption of LTE UL-256QAM by UE manufacturers, which resulted in some RAN vendors omitting UL-256QAM support. Furthermore, due to the hardware limitations of legacy RAN equipment already out in the field, adding the support of UL-256QAM modulation via software update might not always be possible due to hardware limitations, and while MNO can replace the RAN equipment to add UL-256QAM support, it's not financially practical as the number of UE that can take advantage of the functionality is very small.

With the native support of Massive MIMO Active Antenna Unit (AAU) in Time-Division Duplex (TDD) bands, which promise a superior beamforming performance over the existing passive antenna solution \cite{ericsson_2018}, being introduced early in the 5G life cycle, better channel quality with higher SINR is now more common in the real-world environment. Despite the possibility of updating the software of later-generation legacy LTE RAN equipment to support the 5G standard, the performance would be almost identical to the legacy 4G LTE network. In order to obtain the massive capacity and beamforming advantages of 5G, swapping the network equipment for Massive MIMO AAU is required. Additionally, new frequency spectrums dedicated to 5G deployment, particularly the 3.5 GHz or C-Band, were auctioned and allocated around the world, which also requires completely new RAN equipment that is capable of the new frequency band. As a result, many MNOs around the world decided to perform a major overhaul to their network, swapping out their old RAN equipment with the new one that has native 5G support, which usually comes with UL-256QAM-capable hardware \cite{docomo_2019}. Therefore, the adoption of UL-256QAM roses significantly prompting the UE manufacturers to also implement such functionality in their products.

\begin{table}[!tbp]
\setstretch{0.8}
\caption{Summary of SoftBank's 5G Deployment}
\vspace{-1.5mm}
\centering
\label{tab:SoftBankFreqBand}
\resizebox{8.5cm}{!}{\begin{tabular}{@{}lcccc@{}}
\toprule
Frequency Band                  &n3&n28& n77 (3.4G)   & n77 (3.9G)   \\\midrule
Duplex Mode                     &FDD&FDD& TDD       & TDD       \\
Downlink Carrier Frequency (MHz)         &1845-1860& 793-803 & 3400-3440 & 3900-4000       \\
Uplink Carrier Frequency (MHz)         &1750-1765&738-748& 3400-3440 & 3900-4000       \\
Channel Bandwidth (MHz)         &15&10& 40        & 100       \\
Sub Carrier Spacing (SCS)       &15 kHz&15 kHz& 30 kHz    & 30 kHz    \\
TDD Pattern 1 Periodicity (ms) &N/A&N/A& 3 & 3\\
TDD Pattern 1 Slots (DL/UL)     &N/A&N/A& 3/2 & 3/2\\
TDD Pattern 1 Symbols (DL/UL)   &N/A&N/A& 6/4 & 6/4\\
TDD Pattern 2 Periodicity (ms) &N/A&N/A& 2 & 2\\
TDD Pattern 2 Slots (DL/UL)     &N/A&N/A& 4/0 & 4/0\\
TDD Pattern 2 Symbols (DL/UL)   &N/A&N/A& 0/0 & 0/0\\
Maximum Uplink Throughput (Mbps)&90.43&59.52& 55.47 & 142.86\\
\midrule
Dynamic Spectrum Sharing (DSS)&Yes&No&No&No\\
5G (\%)&75.9&61.8&98.7&100.0\\
2T2R (\%)&94.6&96.6&0.0&0.0\\
4T4R (\%)&2.1&0.0&16.7&31.9\\
8T8R (\%)&0.0&0.0&78.0&39.3\\
Massive MIMO (32TRx) (\%)&0.0&0.0&5.2&28.8\\
\bottomrule
\end{tabular}}
\vspace{-7.5mm}
\end{table}

Technically, almost all of the RAN equipment released after the introduction of 5G supports UL-256QAM, but since UL-256QAM can give the MNOs slightly extra capacity with the same spectrum utilization, many RAN vendors charge MNOs for licenses to enable such functionality per cell basis. This led to some MNOs, especially ones in developing nations with tight budgets, being required to strategically enable UL-256QAM on some specific cell or frequency band in order to maximize profit and reduce cost. Therefore, there is a need for the performance evaluations of UL-256QAM to deeply understand the impact it has on live commercial networks so that MNOs around the world can accurately plan their network and avoid unnecessary costs. In this paper, the real-world performance of UL-256QAM on commercial 5G Standalone (SA) networks in two countries; Japan and Thailand; was evaluated. The UE was configured to continuously transmit the data to the server, and then the utilization percentage of each uplink modulation scheme was recorded. The experiments were carried out on both TDD bands commonly found in urban areas and Frequency Division Duplexing (FDD) bands found in rural areas, covering the most common 5G bands used around the world with various mobility characteristics such as walking, driving, taking bus, tram, metro, train, and even high-speed train. Additionally, the impact of UL-256QAM will also be evaluated by modifying the UE modem firmware to disable UL-256QAM support. Finally, the TCP latency will be evaluated with and without UL-256QAM using two TCP congestion control algorithms: Binary Increase Congestion control (BIC) \cite{1354672} and TCP Westwood (TCPW) \cite{10.1145/381677.381704}, in both idle and full load scenario. This paper is organized as follows. Section II will discuss the experiment environment. Section III will provide experimental results as well as an analysis of the results. Finally, the conclusion and future work will be discussed in Section IV.

\vspace{-0.5mm}

\section{Experiment Environment}

\subsection{User Equipment (UE)}

For the UE, Samsung Galaxy S22 Ultra (SC-52C) with Qualcomm Snapdragon X65 5G RF Modem \cite{qualcomm_x65} was used. The UE was verified to support all 5G frequency bands in both countries, including the support for 5G Standalone (SA) on band n3, n28, n41, n77, and n78. It is necessary that the same UE was used throughout the experiment to eliminate the additional variable caused by modem chipset, RF front-end, and antenna design, which vary from one smartphone to another and can have a significant effect on the results as proven in our previous work \cite{10118777}. The modem firmware was modified to toggle UL-256QAM support as needed in each experiment and UECapabilityInformation packet was verified via Network Signal Guru (NSG), a professional network drive test tool, to ensure that the UE has the correct capability before each trial. Additionally, 5G Non-Standalone (NSA) and LTE were turned off to ensure that the UE stays on 5G SA regardless of signal strength. 

\vspace{-1mm}
\subsection{Network Environment}

As for mobile networks, SoftBank Japan and Advanced Info Service (AIS) Thailand were chosen for data collection, as both networks are the only networks in each country that provide nationwide coverage of 5G SA service at the time of study. Commercial SIM cards and plans were used with 5G SA service enabled by each MNO and verified to be working prior to the experiment. Since differences in deployment scheme and enabled feature sets can have a significant impact on the uplink performance, the network overview will be provided to give a clear context prior to the evaluation. 

For SoftBank, 5G service was provided on three frequency bands on four frequency spectra. The summary of all frequency bands operated by SoftBank can be seen in Table \ref{tab:SoftBankFreqBand}. The maximum uplink throughput provided in the table represents the maximum theoretical physical uplink throughput when UL-256QAM is enabled and UL-1Tx is being utilized, which is the maximum capability of the UE used in this study. Furthermore, the deployment information during FY2022 by each frequency spectrum in the Kanto area, where most of the data was collected, as published by the Ministry of Internal Affairs and Communications (Japan) \cite{soumu_2023}, was also included.

\begin{figure}[t!]
\centering
\begin{subfigure}{.24\textwidth}
  \centering
  \includegraphics[width=0.95\linewidth]{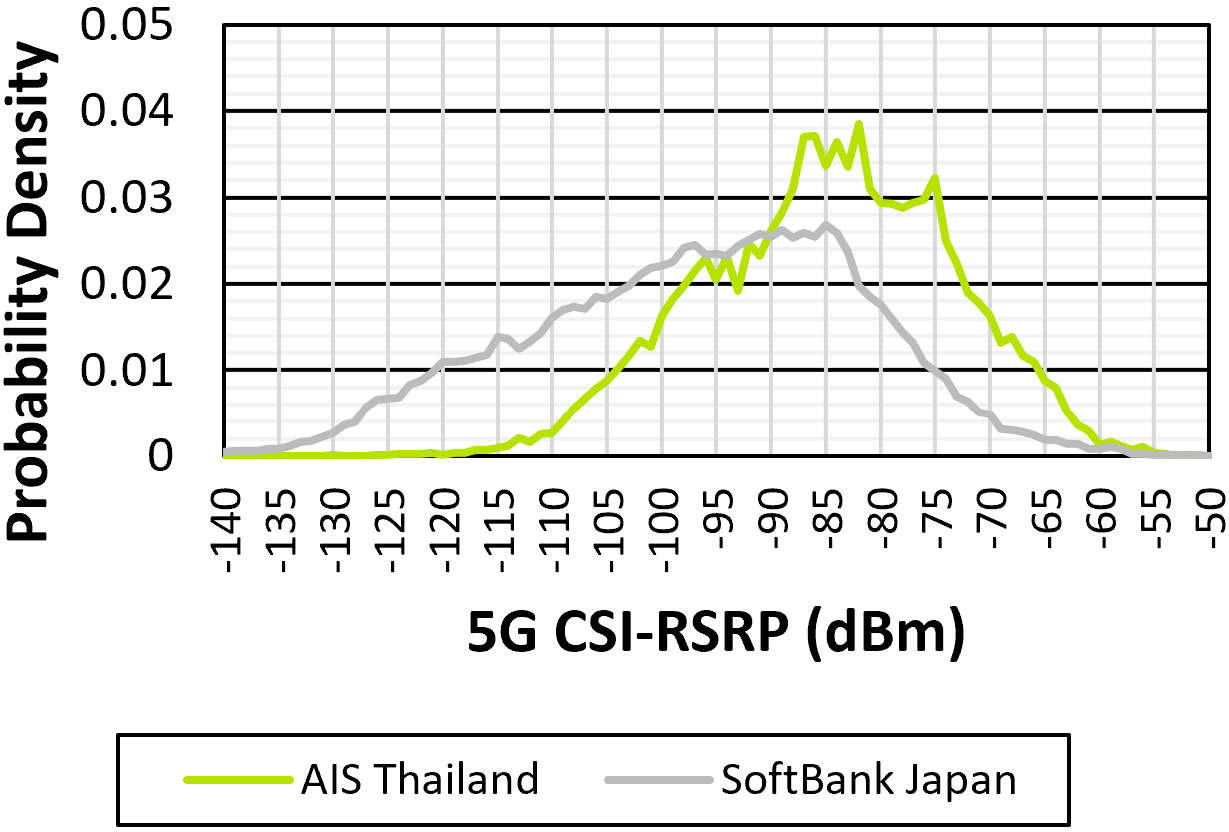}
  \vspace{-1mm}
  \caption{5G CSI-RSRP}
  \label{fig:NetworkRSRPDist}
\end{subfigure}%
\begin{subfigure}{.24\textwidth}
  \centering
  \includegraphics[width=0.95\linewidth]{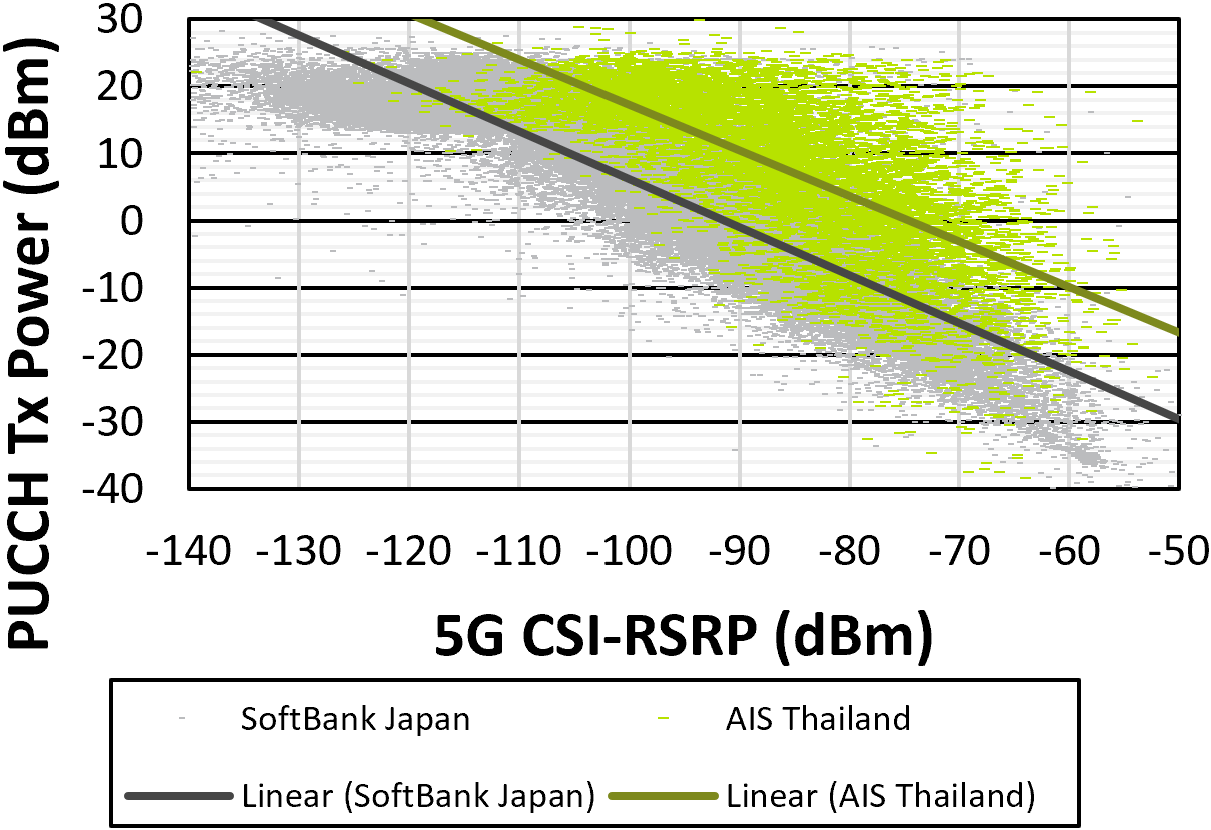}
  \vspace{-1mm}
  \caption{Scatter Plot of 5G CSI-RSRP}
  \label{fig:NetworkRSRPScatter}
\end{subfigure}\\
\begin{subfigure}{.24\textwidth}
  \centering
  \includegraphics[width=0.95\linewidth]{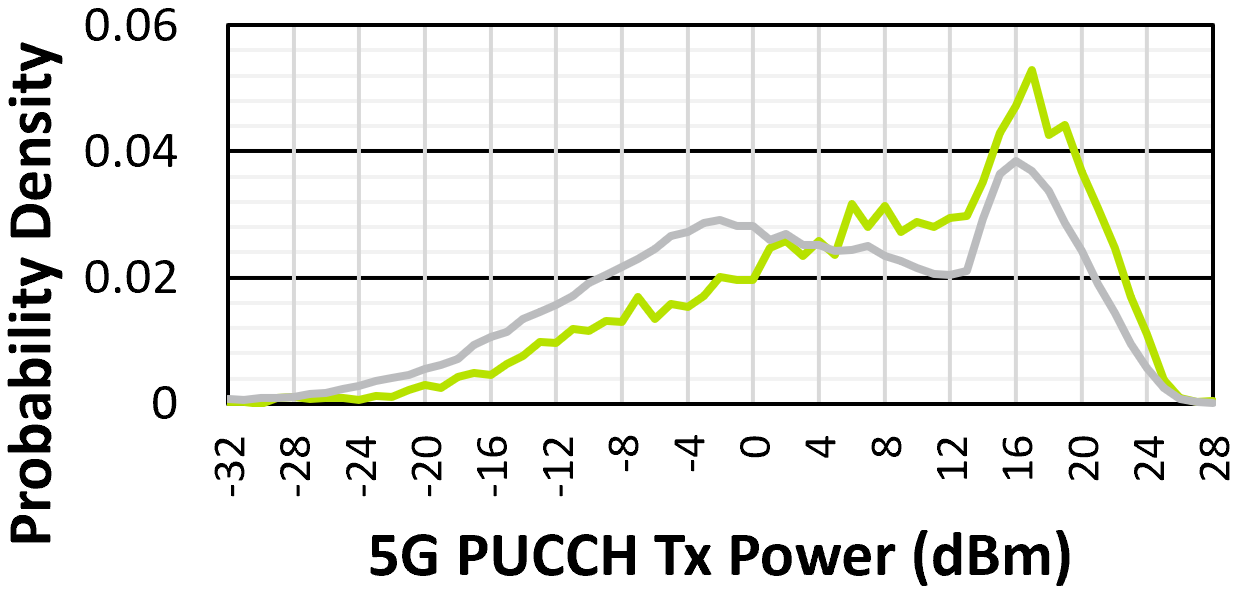}
  \vspace{-1mm}
  \caption{5G PUCCH Tx Power}
  \label{fig:NetworkPUCCHDist}
\end{subfigure}%
\begin{subfigure}{.24\textwidth}
  \centering
  \includegraphics[width=0.95\linewidth]{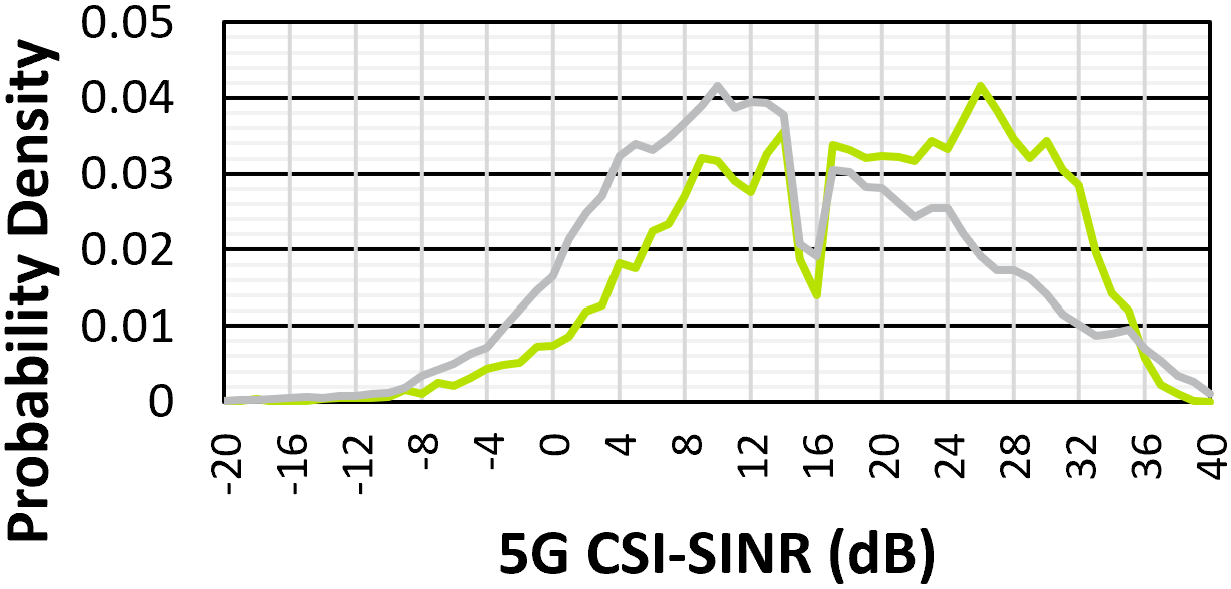}
  \vspace{-1mm}
  \caption{5G CSI-SINR}
  \label{fig:NetworkSINRDist}
\end{subfigure}
\vspace{-1.5mm}
\caption{RF parameters probability distribution by MNO}
\vspace{-3mm}
\label{fig:NetworkDist}
\end{figure}

For AIS, 5G service was provided on three frequency bands on three frequency spectrum. However, 5G band n1 is only deployed at specific tourist attractions for experimental purposes, so it was omitted from this evaluation. As for 5G band n28, AIS utilized a Hybrid DSS approach on this band, which currently has an unsolved connection drop issue when enabling 5G SA. At the time of research, all of the AIS 5G n28 cells were barred from attaching in 5G SA mode, so evaluation on this band couldn't be performed. Therefore, the experiment was only carried out purely on 5G band n41. Unfortunately, the National Broadcasting and Telecommunications Commission (NBTC), doesn't disclose base station information like Japan. However, according to our collected data, AIS utilized 100\% of 64T64R Massive MIMO AAU in the Bangkok area, which should provide an ideal case where beamforming can be fully performed with no interference from conventional passive antenna base stations. On this frequency band, AIS utilizes two channel bandwidths of 60 MHz and 80 MHz, allocating 20-40 MHz of their spectrum to serve LTE depending on cell load in each area. Therefore, the maximum throughput will be given for each channel bandwidth as seen in Table \ref{tab:AISFreqBand}. The throughput for two spectrums that are not being evaluated in this study will also be given for reference purposes.

\begin{table}[!tbp]
\setstretch{0.8}
\caption{Summary of AIS' 5G Frequency Band}
\vspace{-1.5mm}
\centering
\label{tab:AISFreqBand}
\resizebox{8.5cm}{!}{\begin{tabular}{@{}lcccc@{}}
\toprule
Frequency Band                  &n1&n28& n41   & n41   \\\midrule
Duplex Mode                     &FDD&FDD& TDD       & TDD       \\
Downlink Carrier Frequency (MHz)         &2140-2155& 778-798 & 2520-2580 & 2520-2600       \\
Uplink Carrier Frequency (MHz)         &1950-1965&723-743& 2520-2580 & 2520-2600       \\
Channel Bandwidth (MHz)         &15&20& 60        & 80       \\
Sub Carrier Spacing (SCS)       &15 kHz&15 kHz& 30 kHz    & 30 kHz    \\
TDD Pattern 1 Periodicity (ms) &N/A&N/A& 3 & 3\\
TDD Pattern 1 Slots (DL/UL)     &N/A&N/A& 3/2 & 3/2\\
TDD Pattern 1 Symbols (DL/UL)   &N/A&N/A& 6/4 & 6/4\\
TDD Pattern 2 Periodicity (ms) &N/A&N/A& 2 & 2\\
TDD Pattern 2 Slots (DL/UL)     &N/A&N/A& 4/0 & 4/0\\
TDD Pattern 2 Symbols (DL/UL)   &N/A&N/A& 0/0 & 0/0\\
Maximum Uplink Throughput (Mbps)&90.43&116.76& 84.77 & 113.56\\
\bottomrule
\end{tabular}}
\vspace{-7mm}
\end{table}

\vspace{-1.5mm}

\subsection{Test Route and Data Collection}

The data collection was performed via the test function of \textit{Network Signal Guru (NSG)}, a professional network drive test tool, by continuously transmitting data via HTTP POST to the test server and logging all RF parameters in one-second intervals. The test was configured to keep retrying upon disconnection. Typically, UE only reports the highest MCS index that is being used, which doesn't give the full context of the uplink modulation as each resource block may be modulated using a different MCS index. Therefore, the utilization percentage of each modulation scheme is instead being used for evaluation, which will give full information on modulation scheme distribution in each data point. As for the TCP latency test, python implementation of tcp-latency was used. By installing Python and the software directly on the UE, the latency overhead caused by tethering can be eliminated. The tool was configured to transmit a packet every 100 ms and the result was saved to the internal storage.

\begin{figure}[!tbp]
    \centering
    \includegraphics[width=0.91\linewidth]{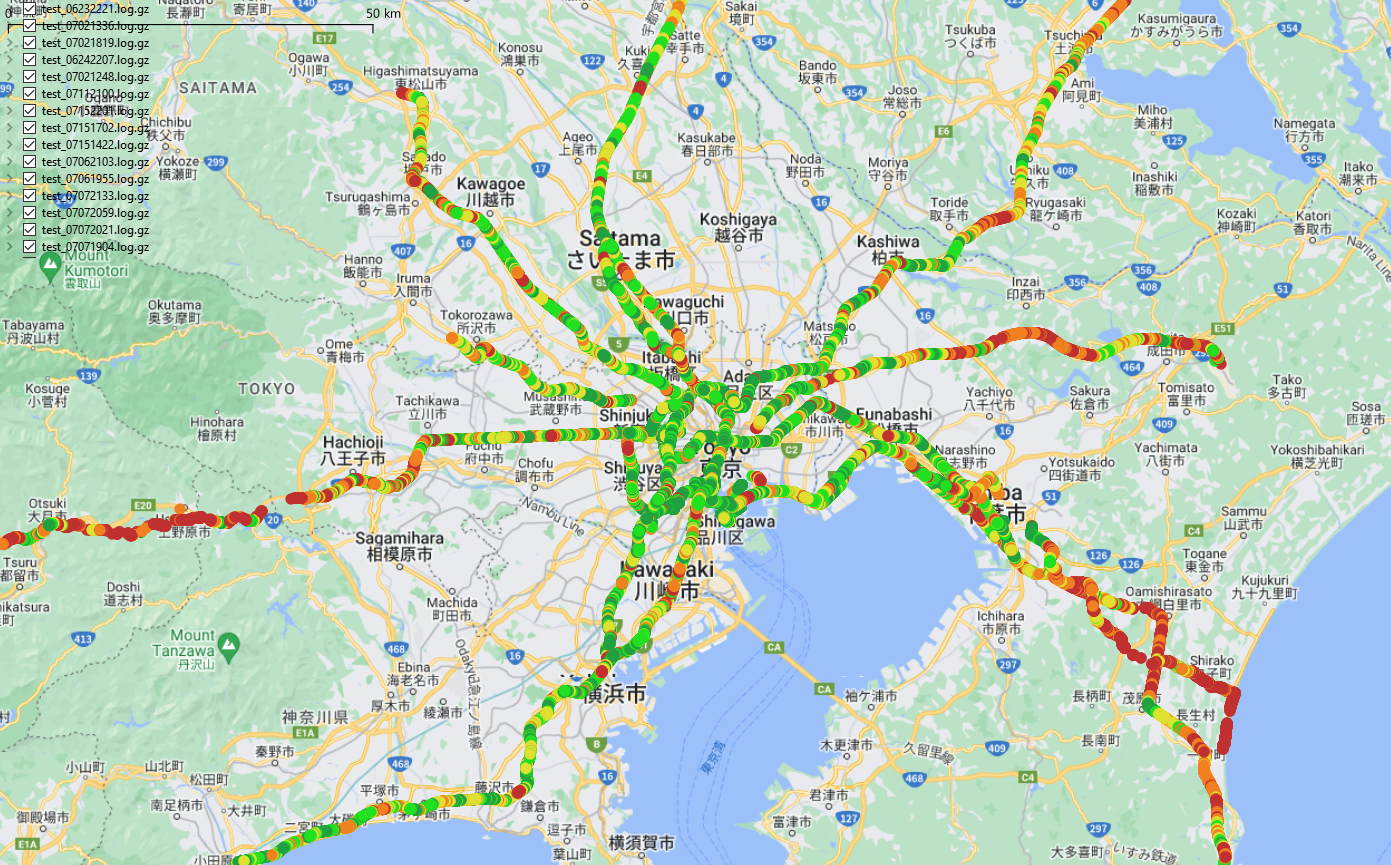}
    \vspace{-1mm}
    \caption{Test Route in Tokyo. Color represents 5G CSI-RSRP.}
    \label{fig:MapTokyo}
    \vspace{-7.5mm}
    
\end{figure}

The data collection was performed on various methods of transportation available in each country. More than 180,000 data points were collected in the time span of more than 50 hours across 58 test routes with more than 300 GB of data being transmitted. It should be noted that some of the AIS base stations didn't have a UL-256QAM license, so the 64QAM MCS Table was assigned to the UE upon attach. The data was manually checked and the data points obtained on the UL-64QAM base station were filtered out before the analysis. The summary of collected data used for evaluation by test case and frequency band can be seen in Table \ref{tab:SummaryData} and \ref{tab:SummaryDataBand}, respective, while the probability distribution of RF parameters of each network can be seen in Fig. \ref{fig:NetworkDist}. Finally, the map of collected data in Tokyo, Japan can be seen in Fig. \ref{fig:MapTokyo}.


\begin{table}[!tbp]
\setstretch{0.8}
\caption{Summary of Evaluation Data by Test Case}
\vspace{-1.5mm}
\centering
\label{tab:SummaryData}
\resizebox{8.7cm}{!}{\begin{tabular}{@{}llccccccc@{}}
\toprule
Country                 &Transportation&Avg.&Data&5G&5G & 5G & 5G &Avg.\\

&Type&Speed&Point&CSI-RSRP& CSI-RSRQ&CSI-SINR&PUCCH&Thpt.\\
&&(km/h)&&(dBm)&(dBm)&(dB)&(dBm)&(Mbps)\\
\midrule
Thailand&Car&56.64&7862&-84.29&-14.38&19.56&7.85&33.99\\
&Train&45.54&3104&-86.62&-14.73&15.30&7.21&28.46\\
\midrule
\multicolumn{2}{c}{Thailand Average}&53.47&10966&-84.95&-14.48&18.35&7.65&32.43\\
\midrule
Japan&High-Speed Rail&189.10&12649&-98.72&-16.40&13.39&4.68&15.34\\
&Rural Train&65.04&49466&-99.67&-15.91&14.17&5.34&13.69\\
&Suburban Train&54.54&29053&-96.23&-15.98&13.09&3.17&13.91\\
&Car&33.21&11567&-92.58&-15.78&13.61&2.03&20.45\\
&Elevated Metro&25.86&3636&-83.31&-16.16&10.52&-6.91&31.13\\
&Bus/Tram&14.59&6522&-92.83&-15.07&17.61&-0.68&21.18\\
&Walk&3.29&1561&-83.83&-15.56&14.74&-8.54&39.21\\
\midrule
\multicolumn{2}{c}{Japan Average}&68.27&114454&-96.85&-15.92&13.84&3.44&15.95
\\
\bottomrule

\end{tabular}}
\vspace{-3mm}
\end{table}

\begin{table}[!tbp]
\caption{Summary of Evaluation Data by Band}
\setstretch{0.8}
\vspace{-1.5mm}
\centering
\label{tab:SummaryDataBand}
\resizebox{8.7cm}{!}{\begin{tabular}{@{}lcccccccc@{}}
\toprule
Country                 &Band&B/W&Data&5G&5G & 5G & 5G &Avg.\\

&&(MHz)&Point&CSI-RSRP& CSI-RSRQ&CSI-SINR&PUCCH&Thpt.\\
&&&&(dBm)&(dBm)&(dB)&(dBm)&(Mbps)\\
\midrule
Thailand&n41&60&10433&-85.12&-14.49&18.36&7.82&32.02\\
&n41&80&533&-81.54&-14.38&18.26&3.44&40.28\\
\midrule
\multicolumn{3}{c}{Thailand Average}&10966&-84.95&-14.48&18.35&7.65&32.43\\
\midrule
Japan&n3&15&44253&-92.42&-16.08&15.32&1.13&20.15\\
&n28&10&22509&-99.48&-16.12&11.32&3.80&15.06\\
&n77 (3.4G)&40&38885&-99.26&-15.89&12.86&6.13&12.42\\
&n77 (3.9G)&100&8807&-101.67&-14.79&17.23&5.58&25.75\\
\midrule
\multicolumn{3}{c}{Japan Average}&114454&-96.85&-15.92&13.84&3.44&15.95\\
\bottomrule

\end{tabular}}
\vspace{-4mm}
\end{table}

While the best RF parameters to use for the evaluation are the RSRP and SINR at the base station side, such data is only obtainable by the MNOs. On the UE side, Timing Advance (TA), which is used for subframe synchronization, can give a rough distance of the path that RF takes to reach the base station, but this parameter is only updated during the initial attach or handover, which doesn't happen very often. While a moderate correlation was found between Downlink SS-RSRP and Uplink modulation in the previous study, SS-RSRP is only suitable for comparison of the same network due to the difference in transmission power configured at the base station of each MNO. As seen in Fig. \ref{fig:NetworkRSRPDist} and Fig. \ref{fig:NetworkRSRPScatter}, the AIS base station outputs about 15 dB higher power compared to SoftBank. Hence, PUCCH Tx Power and Downlink CSI-SINR will be used for the evaluation. PUCCH Tx Power represents the power that the UE takes to transmit control information and signaling without accounting for the power takes to transmit user data and typically the UE will raise this power depending on the difficulty it has to reach the base station until it hits the limit of 23 dBm for non-HPUE bands. Therefore, while this parameter is affected by the type of antenna used on the base station, it's still an appropriate parameter to help determine the communication difficulty between UE and the base station. According to the probability density of PUCCH Tx Power (see Fig. \ref{fig:NetworkPUCCHDist}), the distribution on the two networks is similar, showing that UE has a similar level of difficulty establishing the connection with base stations. On the other hand, CSI-SINR represents Signal to Interference \& Noise Ratio for the downlink, which can be used to infer communication channel and cell planning quality. From Fig. \ref{fig:NetworkSINRDist}, it is clear that AIS deployment of 100\% Massive MIMO AAU gives them a significant SINR advantage over SoftBank's mixed deployment, and delivers less neighbor interference and better channel quality, which is an ideal condition for UL-256QAM.

\vspace{-2mm}

\section{Results and Analysis}

\subsection{Utilization of UL-256QAM}

\begin{table}[!tbp]
\caption{Percentage of Modulation Utilization by Test Case}
\setstretch{0.8}
\vspace{-1.5mm}
\centering
\label{tab:ModUtilization}
\resizebox{8.7cm}{!}{\begin{tabular}{@{}llcccccc@{}}
\toprule
Country                 &Transportation&Avg.&QPSK&16QAM & 64QAM & 256QAM &Avg.\\

&Type&Speed&(\%)&(\%)&(\%)&(\%)&Bit/\\
&&(km/h)&&&&&Symbol\\
\midrule
Thailand&Car&56.64&10.84&11.78&28.74&48.63&6.30\\
&Train&45.54&10.37&11.43&26.87&51.32&6.38\\
\midrule
Japan&High-Speed Rail&189.10&34.80&17.70&33.79&13.71&4.53\\
&Rural Train&65.04&36.62&16.27&29.94&17.16&4.55\\
&Suburban Train&54.54&28.40&18.77&34.20&18.63&4.86\\
&Car&33.21&22.32&18.39&40.14&19.15&5.12\\
&Elevated Metro&25.86&9.88&8.81&33.64&47.67&6.38\\
&Bus/Tram&14.59&23.02&13.33&41.82&21.83&5.25\\
&Walk&3.29&0.61&9.24&39.18&50.97&6.81\\

\bottomrule
\end{tabular}}
\vspace{-3mm}
\end{table}

\begin{table}[!tbp]
\caption{Percentage of Modulation Utilization by Band}
\setstretch{0.8}
\vspace{-1.5mm}
\centering
\label{tab:ModUtilizationBand}
\resizebox{8cm}{!}{\begin{tabular}{@{}lcccccc@{}}
\toprule
Band                 &Massive&QPSK&16QAM & 64QAM & 256QAM &Avg.\\

&MIMO&(\%)&(\%)&(\%)&(\%)&Bit/\\
&(\%)&&&&&Symbol\\
\midrule
n3 (FDD 1.8 GHz)&0.0&25.26&17.42&38.98&18.34&5.01\\
n28 (FDD 700 MHz)&0.0&30.66&16.64&31.32&21.38&4.87\\
n41 (TDD 2.6 GHz)&100.0&10.79&11.84&28.20&49.17&6.32\\
n77 (TDD 3.4 GHz)&5.2&34.08&16.68&28.66&20.57&4.71\\
n77 (TDD 3.9 GHz)&28.8&46.43&14.13&29.36&10.08&4.06\\

\bottomrule
\end{tabular}}
\vspace{-7mm}
\end{table}

\looseness=-1
From the experiment results (see Table \ref{tab:ModUtilization}), it has been found that as the mobility speed increased, the utilization of UL-256QAM decreased. SoftBank Japan's heavy uses of passive antenna and RRU resulted in sub-par beamforming performance causing the base station to be unable to track the UE very accurately, which ended up with UL-256QAM being under-utilized and most of the resource blocks were modulated using 64QAM when mobility speed was beyond 40 km/h. When looking at spectral efficiency, it was found that anything faster than a car resulted in a spectral efficiency of less than 5 bits/symbol, a number achievable with UL-64QAM. However, when considering the AIS Thailand's cases, the results changed dramatically, the use of 100\% 64TRx Massive MIMO AAU resulted in very accurate tracking of UE. Even though the UE is moving at the speed of more than 55 km/h, almost half of the resource blocks were modulated using 256QAM, which causes spectral efficiency to be well over 6 bits/symbol, resulting in UL-256QAM delivering a significantly improved uplink performance compared to UL-64QAM.

\begin{figure}[t!]
\centering
\begin{subfigure}{.24\textwidth}
  \centering
  \includegraphics[width=0.95\linewidth]{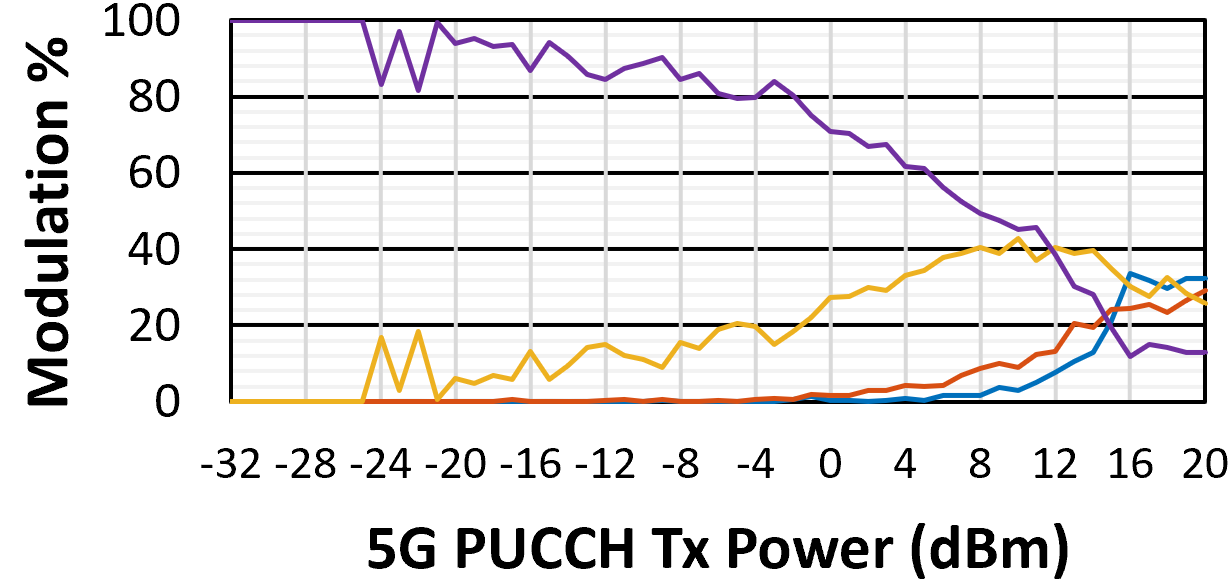}
  \vspace{-1mm}
  \caption{AIS Modulation vs. PUCCH}
  \label{fig:AISPUCCHModulation}
\end{subfigure}%
\begin{subfigure}{.24\textwidth}
  \centering
  \includegraphics[width=0.95\linewidth]{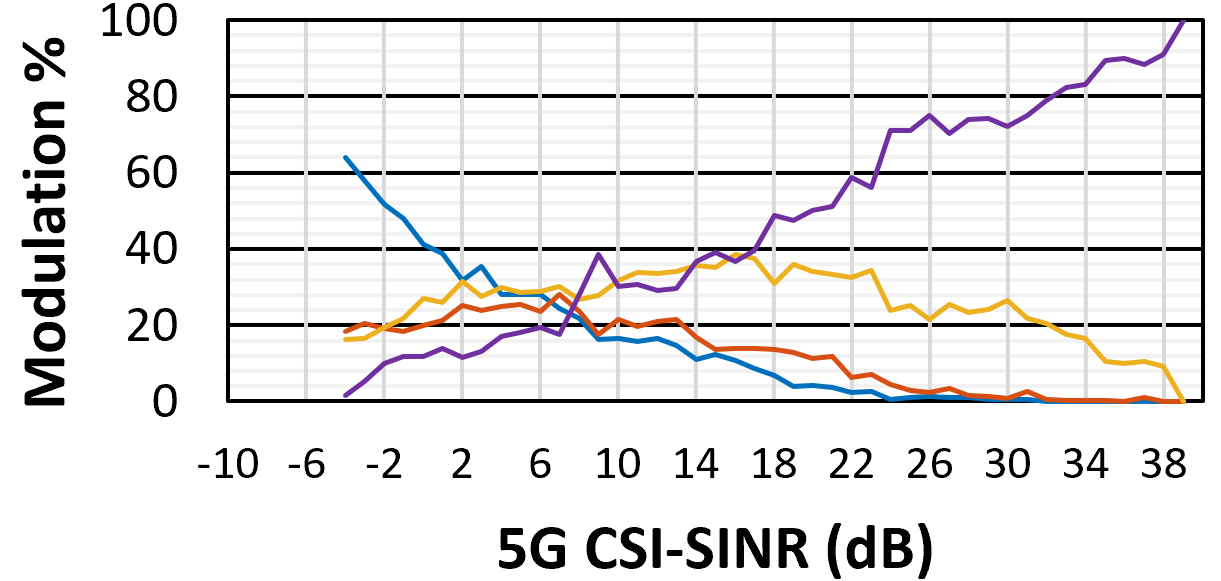}
  \vspace{-1mm}
  \caption{AIS Modulation vs. CSI-SINR}
  \label{fig:AISSINRModulation}
\end{subfigure}\\
\begin{subfigure}{.24\textwidth}
  \centering
  \includegraphics[width=0.95\linewidth]{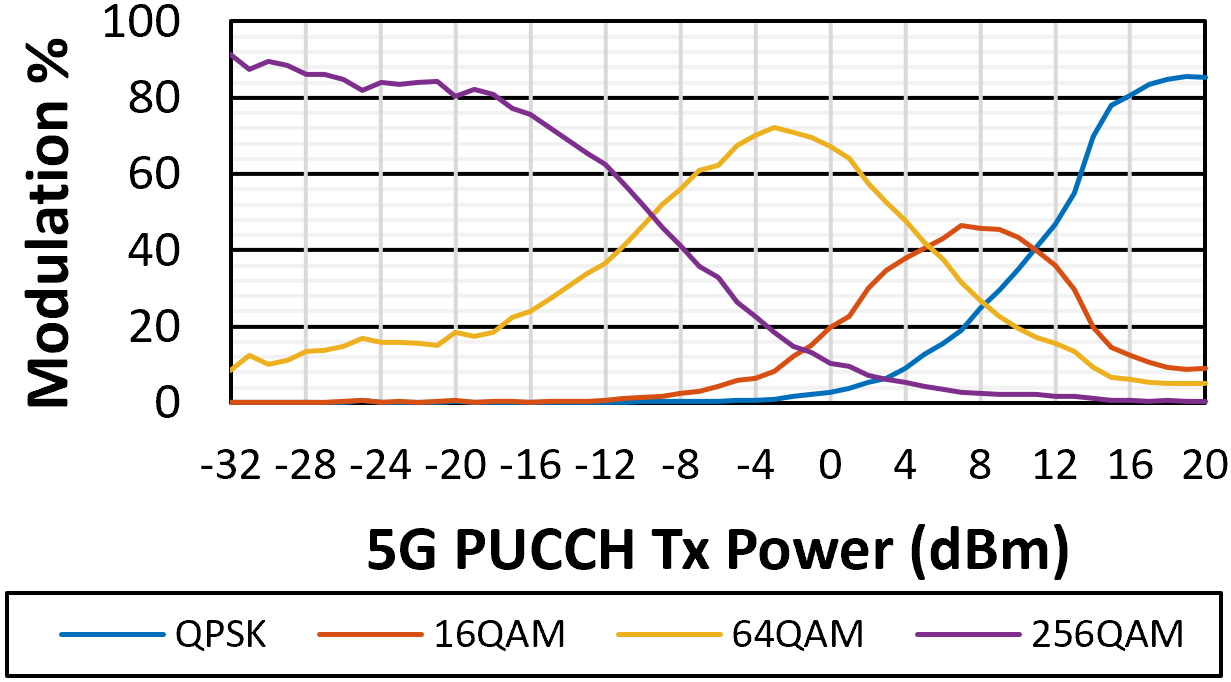}
  \vspace{-1mm}
  \caption{SB Modulation vs. PUCCH}
  \label{fig:SoftBankPUCCHModulation}
\end{subfigure}%
\begin{subfigure}{.24\textwidth}
  \centering
  \includegraphics[width=0.95\linewidth]{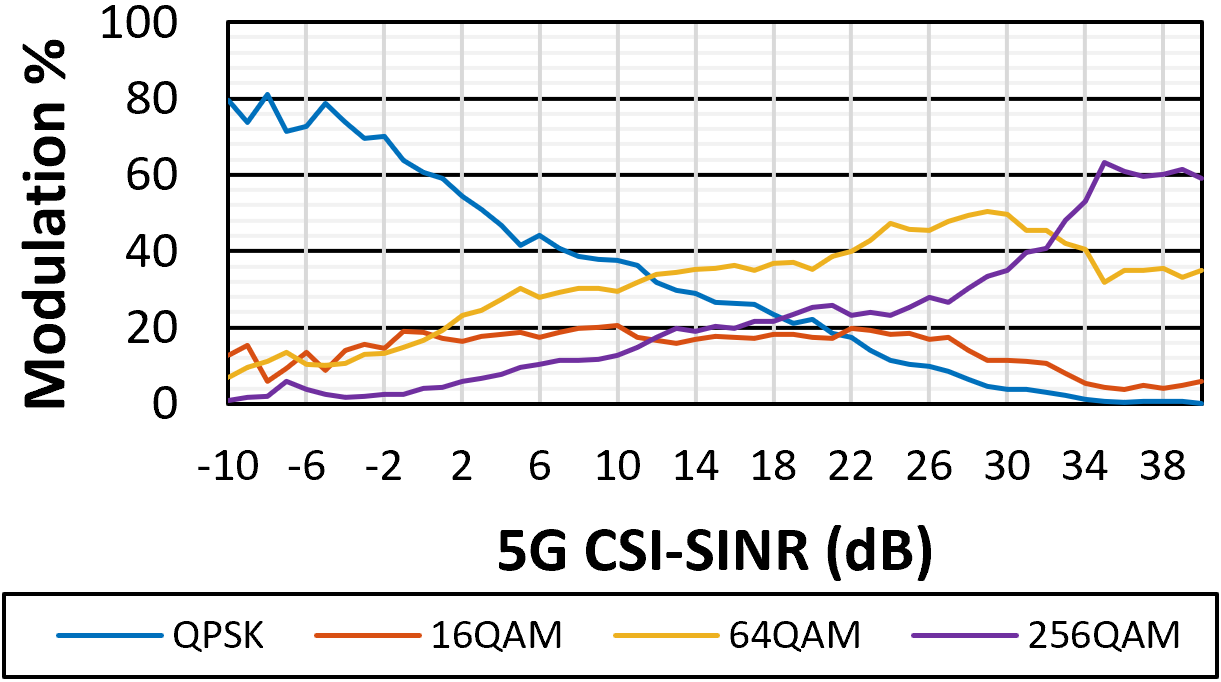}
  \vspace{-1mm}
  \caption{SB Modulation vs. CSI-SINR}
  \label{fig:SoftBankSINRModulation}
\end{subfigure}
\begin{subfigure}{.24\textwidth}
  \centering
  \includegraphics[width=0.95\linewidth]{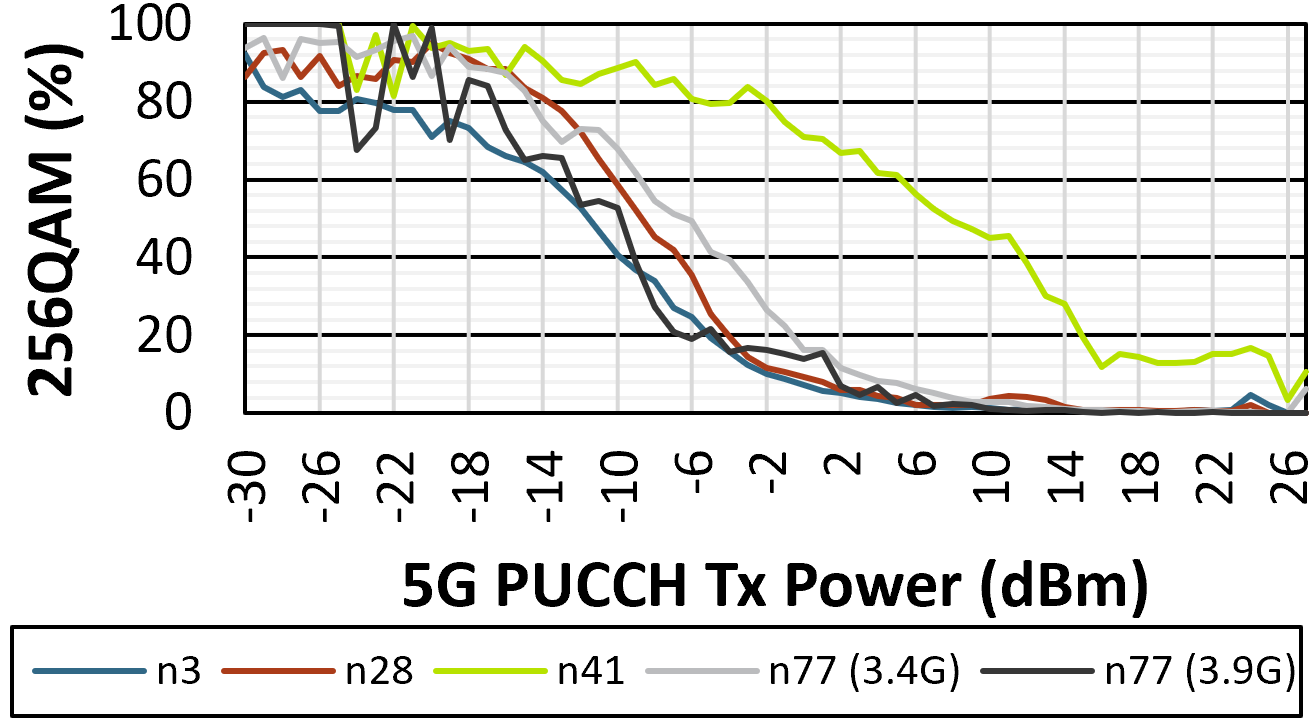}
  \captionsetup{justification=centering}
  \vspace{-1mm}
  \caption{256QAM Utilization (\%) \\ vs. PUCCH Tx Power per Band}
  \label{fig:ModulationBandPUCCH}
\end{subfigure}%
\begin{subfigure}{.24\textwidth}
  \centering
  \includegraphics[width=0.95\linewidth]{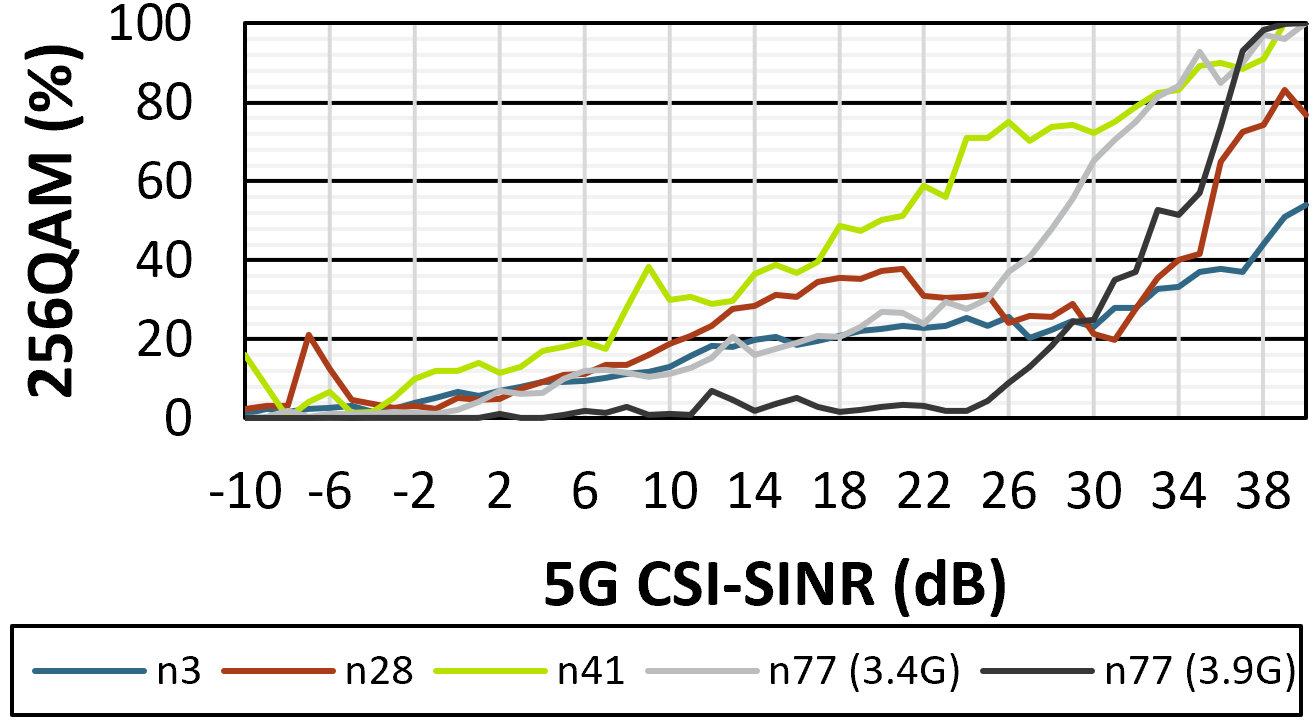}
  \captionsetup{justification=centering}
  \vspace{-1mm}
  \caption{256QAM Utilization (\%) \\ vs. CSI-SINR per Band}
  \label{fig:ModulationBandSINR}
\end{subfigure}\\
\vspace{-2mm}
\caption{Modulation Utilization vs RF Parameters}
\label{fig:ModulationUtilization}
\vspace{-9.5mm}
\end{figure}

Furthermore, when looking at the modulation utilization by frequency band (see Table \ref{tab:ModUtilizationBand}), it has been found that as the frequency increased, the 256QAM utilization decreased, which likely due the result of higher path loss at higher carrier frequency reducing the uplink signal quality, resulting in slightly higher spectral efficiency on lower frequency FDD bands like n3 and n28 compared to higher frequency TDD bands like n77 as more resource blocks were able to be modulated using 256QAM. However, even when considering the low-frequency FDD band like n28, the spectral efficiency was still less than 5 bits/symbol with 256QAM utilization of 21.38\%. On the other hand, the full deployment of Massive MIMO by AIS paid off in a big way as 49.17\% of the resource blocks were modulated using 256QAM driving the spectral efficiency to 6.32 bits/symbol, well beyond the limit of UL-64QAM and even outperform lower frequency bands counterpart by a significant margin, despite using the carrier frequency of 2.6 GHz. According to 3GPP's studies \cite{Simulation_1}\cite{Simulation_2}\cite{Simulation_3}, the simulation results show that the SNR requirement to reach 70\% of the maximum throughput at the carrier frequency of 4.0 GHz, channel bandwidth of 40 MHz, and same the TDD timing with commercial networks in this paper, when using UL-256QAM at MCS Index of 24 on UL-1Tx UE with 2T2R, 4T4R, and 8T8R base stations are 21.29, 17.23, and 13.87 dB, respectively. Furthermore, it was found that channel bandwidth of 100 MHz requires about 1.5 dB higher SNR to reach 70\% of the maximum throughput compared to the 10 MHz counterpart. \looseness=-1

While the numbers for 16Rx, 32Rx, and 64Rx were not included in the 3GPP studies, it can be seen from the simulation results that doubling the number of Rx at the base station side yields about 3.36 to 4.16 dB reduction in the SNR requirement without channel impairment and 3.43 to 4.09 dB reduction with channel impairment. The improvement becomes smaller as more receiving antennae are added at the base station. Therefore, by extrapolating the known data, the 64TRx Massive MIMO base station should yield an approximate 7.5-9.5 dB reduction in the SNR requirement over 8T8R. Because SoftBank deploys a mix of 16.7\% 4T4R and 78.0\% 8T8R on frequency band n77 (3.4 GHz), the actual number should be closer to 8T8R than 4T4R. As the actual uplink SNR at the base station can't easily be obtained, the difference of downlink SINR to reach UL-256QAM utilization of 50\% will be used. It was found that on SoftBank's band n77 (3.4 GHz), SINR of 28.5 dB was required to hit 50\% UL-256QAM utilization, whereas SINR of 20 dB is required to hit the target on AIS' band n41 (2.6 GHz). This shows an approximate reduction of SNR requirement by 8.5 dB, which agrees with the 3GPP's studies. Because the manufacturing of low-frequency band Massive MIMO is impractical due to long-wavelength requiring large antenna elements, it can be said that UL-256QAM will only improve capacity and throughput by a significant margin when used alongside a full Massive MIMO deployment on middle-frequency bands.

It should be noted that 3GPP did propose a High Power UE (HPUE) standard for middle-frequency TDD band: n41/n77/n78/n79 \cite{3GPP_38-101-1}, and AIS did allow HPUE Power Class 2 (PC2) on their n41 band for the total power budget of 26 dBm, but SoftBank didn't allow HPUE operation at the time of research. Hence, to keep things fair, a Japanese UE with no PC2 capability was used throughout the research.
\begin{table}[!tbp]
\captionsetup{justification=centering}
\caption{Uplink Transmission comparison with\\ different PUSCH MCS Tables}
\setstretch{0.8}
\vspace{-1mm}
\centering
\label{tab:PUSCHCompared}
\resizebox{8.7cm}{!}{\begin{tabular}{@{}lcccccccc@{}}
\toprule
Band                 &Case&TCP Congestion&MCS&5G&5G& 64QAM & 256QAM &Scheduled\\

&&Control&Table&CSI-SINR&PUCCH&(\%)&(\%)&Resource\\
&&Algorithm&&(dB)&(dBm)&&&Block (\%)\\
\midrule
n3&Rural Train&BIC&qam64&10.44&4.54&45.37&0.00&34.99\\
&&BIC&qam256&9.93&5.15&29.67&10.22&27.65\\
\midrule
n28&Rural Train&BIC&qam64&7.14&6.09&46.96&0.00&54.71\\
&&BIC&qam256&7.84&5.52&32.59&19.11&56.55\\
\midrule
n28&Urban Train&BIC&qam64&11.02&-3.45&75.20&0.00&63.00\\
&&BIC&qam256&10.90&-1.82&42.49&25.70&65.38\\
&&TCPW&qam64&10.77&-2.20&74.18&0.00&59.33\\
&&TCPW&qam256&10.58&-1.32&47.45&28.42&62.17\\

\midrule
n77&Urban Train&BIC&qam64&13.22&5.80&49.56&0.00&38.43\\
&&BIC&qam256&11.67&6.69&25.00&20.34&47.92\\
&&TCPW&qam64&12.96&5.29&46.30&0.00&41.31\\
&&TCPW&qam256&13.23&8.49&28.26&12.19&40.74\\

\bottomrule
\end{tabular}}
\vspace{-4.5mm}
\end{table}

\begin{figure}[t!]
\centering
\begin{subfigure}{.24\textwidth}
  \centering
  \includegraphics[width=0.95\linewidth]{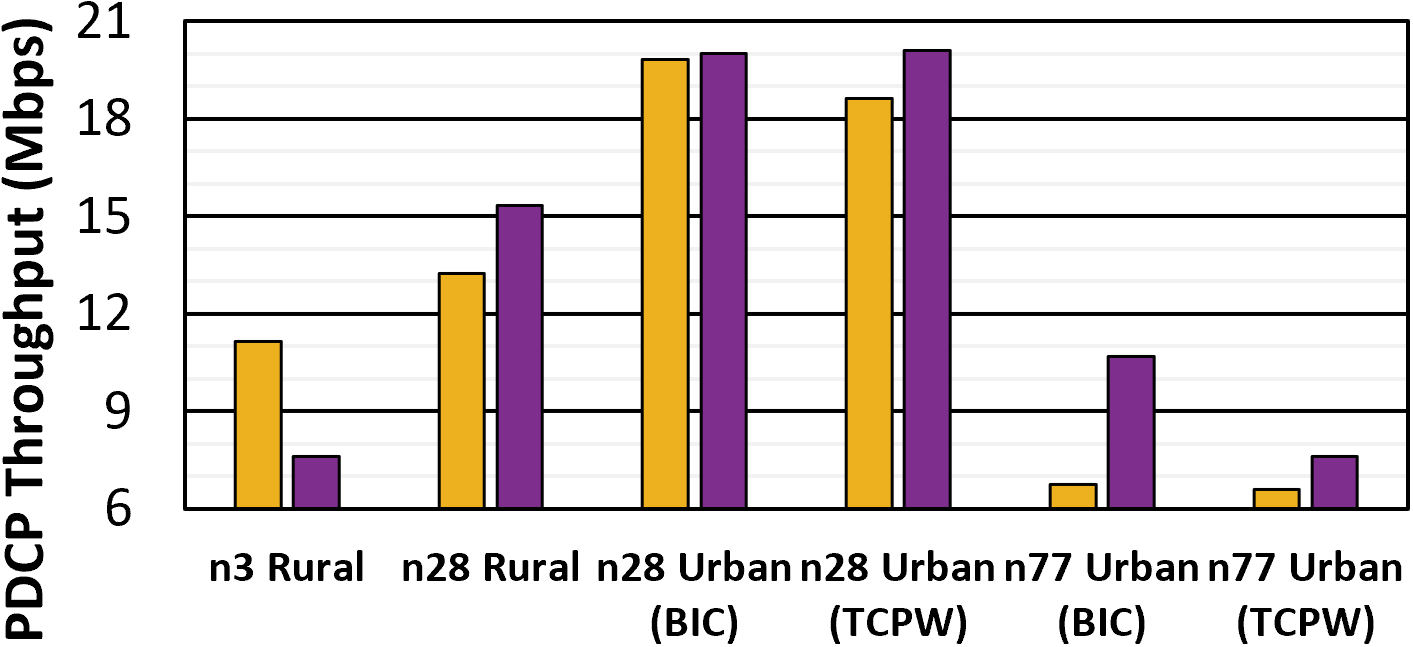}
  \vspace{-1mm}
  \caption{PDCP Throughput}
  \label{fig:ThroughputRaw}
\end{subfigure}%
\begin{subfigure}{.24\textwidth}
  \centering
  \includegraphics[width=0.95\linewidth]{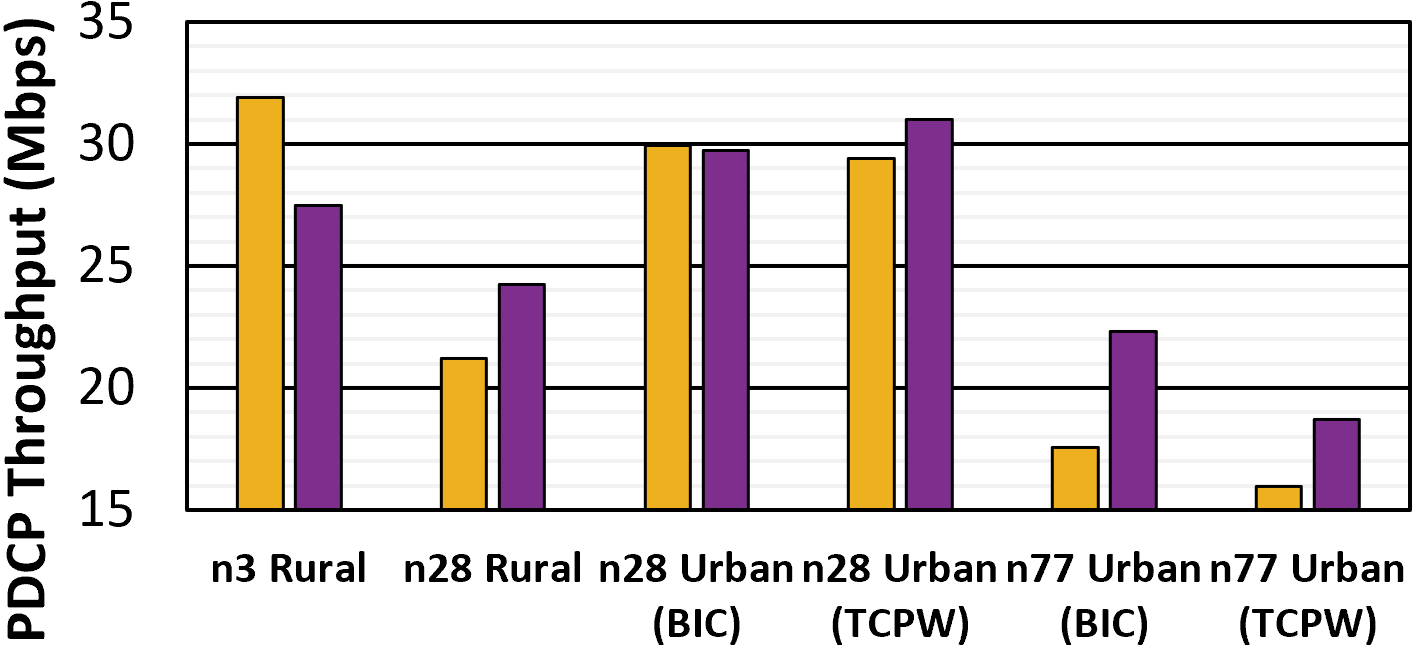}
  \vspace{-1mm}
  \caption{Normalized PDCP Thpt.}
  \label{fig:ThroughputNormalized}
\end{subfigure}\\
\begin{subfigure}{.24\textwidth}
  \centering
  \includegraphics[width=0.95\linewidth]{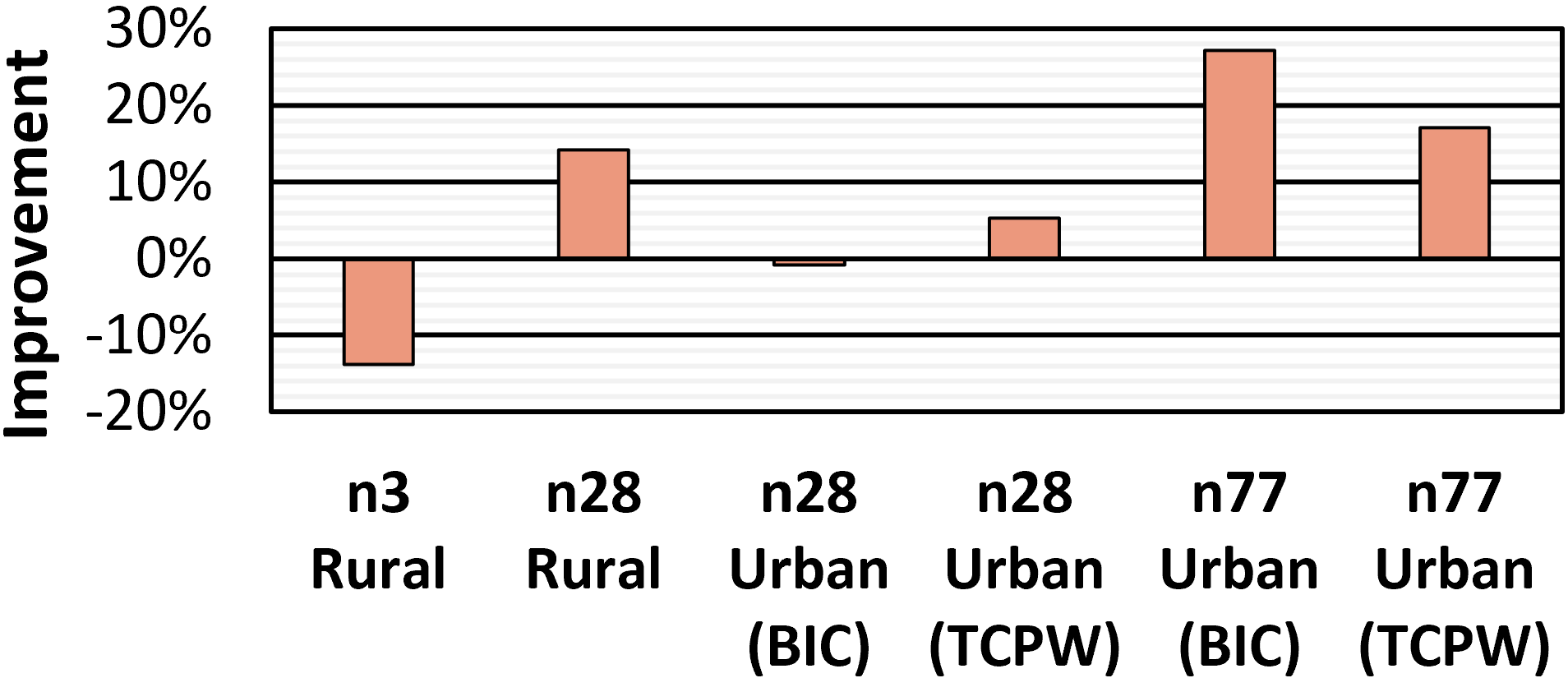}
  \vspace{-1mm}
  \caption{256QAM Improvement (\%)}
  \label{fig:ComparisonNormalized}
\end{subfigure}%
\begin{subfigure}{.24\textwidth}
  \centering
  \includegraphics[width=0.95\linewidth]{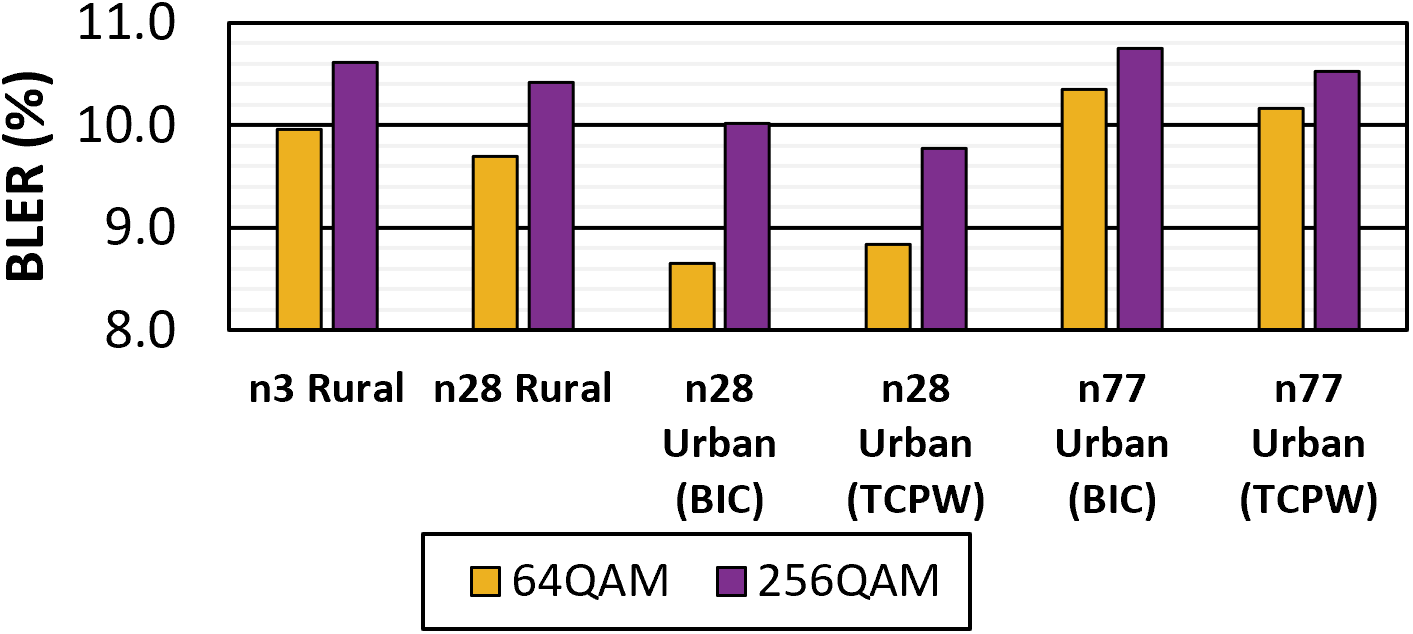}
  \vspace{-1mm}
  \caption{Block Error Rate (\%)}
  \label{fig:BLER}
\end{subfigure}
\vspace{-1.8mm}
\caption{Results from Throughput Experiment}
\vspace{-8mm}
\label{fig:ThptTest}
\end{figure}

When looking at Modulation Percentage vs. PUCCH (see Fig. \ref{fig:AISPUCCHModulation} and Fig. \ref{fig:SoftBankPUCCHModulation}), which shows the utilization percentage of each modulation scheme against the control channel transmission power used by UE. Typically, UE will raise the transmission power at a longer distance to reach the base station. It has been found that UL-256QAM was always active on AIS' full Massive MIMO network including at cell edges. However, on SoftBank, 256QAM utilization falls below 64QAM modulation very quickly at PUCCH Tx Power of -10 dBm and became completely inactive when the power is above 8 dBm, which is typical for indoor usage, showing that UL-256QAM provided only a slight benefit for SoftBank. Furthermore, when looking at Modulation Percentage vs CSI-SINR (see Fig. \ref{fig:AISSINRModulation} and \ref{fig:SoftBankSINRModulation}), where higher CSI-SINR represents lower downlink interference from neighbor cells, commonly found at the cell center. With full Massive MIMO deployment, when the UE is at the cell center with a line of sight to the base station, more than 90\% of the resources block were guaranteed to be modulated in 256QAM, giving a massive uplink throughput boost. Unfortunately, even at the cell center, SoftBank struggled to maintain good utilization of 256QAM resulting in only a little improvement over UL-64QAM. From per-band data (see Fig. \ref{fig:ModulationBandPUCCH} and Fig. \ref{fig:ModulationBandSINR}), AIS' n41 can be seen outperforming all of the SoftBank bands, demonstrating a clear advantage in favor of Massive MIMO AAU.
\vspace{-2mm}
\subsection{Uplink Throughput with Passive Antenna Deployment}
\vspace{-0.5mm}
\begin{figure}[!tbp]
    \centering
    \includegraphics[width=0.85\linewidth]{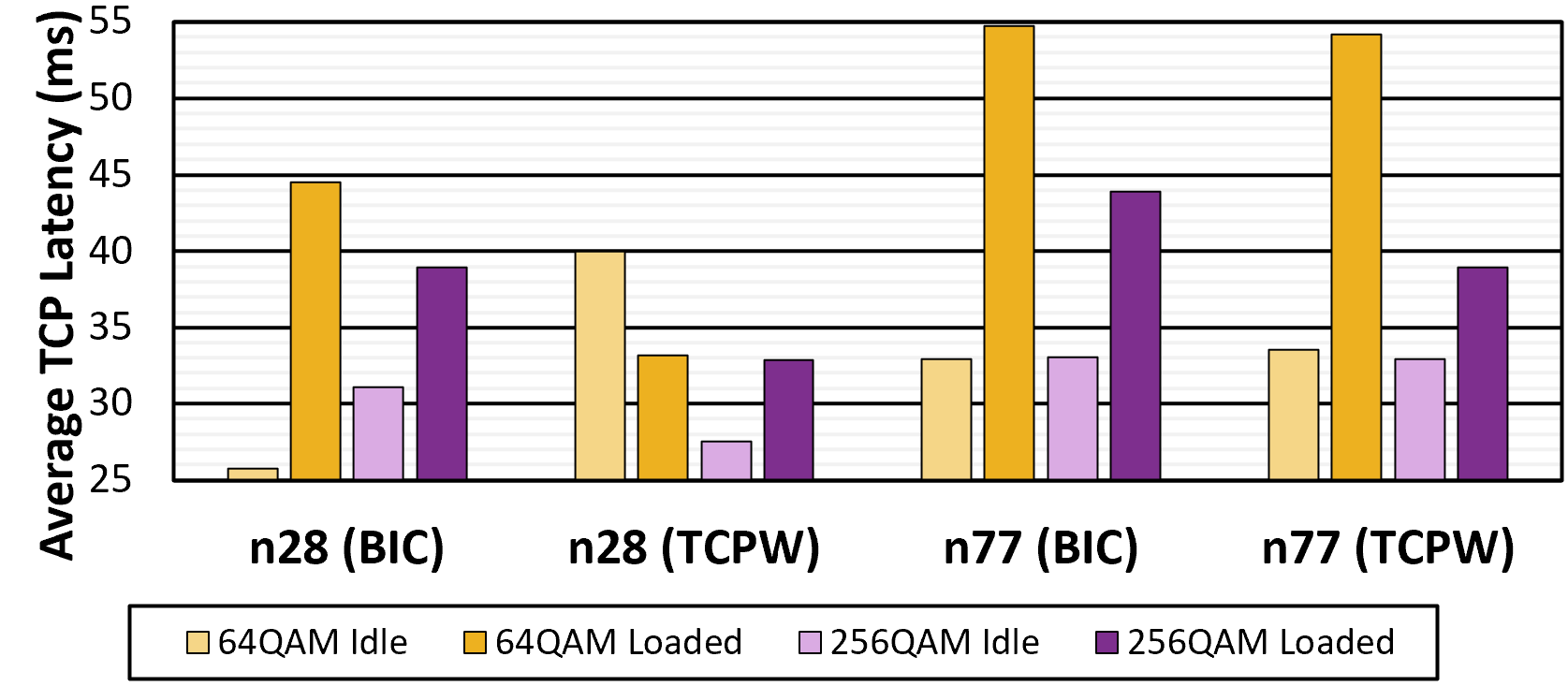}
    \vspace{-1.5mm}
    \caption{Average TCP Latency in both idle and full load scenarios on two frequency bands: n28 (FDD) and n77 (TDD) with two TCP congestion control algorithms: BIC and TCPW.}
    
    \label{fig:TCPLatency}
    \vspace{-7.7mm}
\end{figure}

Due to the inconsistency of transportation in Thailand, a fair comparison of uplink throughput between UL-64QAM and UL-256QAM MCS Table couldn't be conducted. In Japan, train punctuality is one of the world's best, so the experiment was conducted on the same route with different PUSCH MCS tables pre-configured in the modem's firmware. A frequency band lock was used to force UE on the desired band. The Rural Train experiments were conducted on the JR Musashino Line running through Saitama and Chiba prefecture and the Urban Train experiments were conducted by riding F Liner Limited Express on the Tokyu Toyoko Line running from Shibuya to Yokohama. The UE was placed on the train window facing the same side on the same carriage in all experiments to ensure consistency in result collection. The measured throughput is provided in Fig. \ref{fig:ThroughputRaw}, but since the cell load varies depending on users' activities, the number of resource block scheduled was recorded (see Table \ref{tab:PUSCHCompared}), and all throughput result was normalized to reflect the throughput when UE has access to all of the resource on each frequency band (see Fig. \ref{fig:ThroughputNormalized}). 

From the results on rural trains, it has been found that using the UL-256QAM MCS Table increased the throughput by 14.17\% on band n28. However, on Dynamic Spectrum Sharing (DSS) enabled band n3, enabling UL-256QAM resulted in a 13.79\% uplink throughput decline as seen in Fig. \ref{fig:ComparisonNormalized}. When considering urban train cases, where two TCP congestion algorithms were evaluated, it has been found that UL-256QAM provided almost negligible throughput improvement on band n28 with BIC TCP, but 5.33\% throughput uplift was observed when using TCPW. When comparing BIC to TCPW with UL-256QAM enabled, it was found that TCPW delivers 4.29\% higher throughput on FDD bands. On the other hand, on TDD band n77, it was found that UL-256QAM did, in fact, deliver 27.18\% and 17.16\% higher throughput on BIC TCP and TCPW, respectively, with BIC TCP providing 16.78\% higher throughput than TCPW. When Block Error Rate (BLER) is considered (see Fig. \ref{fig:BLER}), the results show that UL-256QAM increased the BLER across the board. When comparing to UL-64QAM, BLER on band n28 with UL-256QAM enabled when riding urban trains was increased by an average of 1.15\% compared to 0.72\% on rural trains. Since higher BLER resulted in more errors that has to be corrected, increasing the overhead, this explains the negligible gain in throughput observed in this experiment. Finally, the average BLER increase on band n77 was 0.39\%, which is why the largest difference in throughput was observed as the extra bits were utilized to transmit the user's data instead of error correction.

\vspace{-1.5mm}
\subsection{TCP Latency}

The TCP latency results, as shown in Fig. \ref{fig:TCPLatency}, demonstrated that in the idle scenario, despite one outlier, FDD band did deliver lower latency than TDD counterpart with the largest difference of 5.37 ms observed when using TCPW with UL-256QAM enabled. In idle scenarios, it was found that both TCP congestion control algorithms delivering a comparable performance. However, when the link is fully loaded, TCPW delivers between 0.51 ms and 11.36 ms with an average of 5.73 ms lower latency compared to BIC TCP. Finally, with UL-256QAM enabled, the latency performance in loaded scenarios was improved across the board with an improvement of between 0.28 ms and 15.24 ms with an average of 7.97 ms observed on urban trains.
\vspace{-1mm}
\section{Conclusions and Future Work}

In this paper, the impact of Uplink 256QAM (UL-256QAM) was evaluated on commercial 5G Standalone (SA) networks in two countries with major differences in deployment schemes on various types of transportation and mobility profiles. It was found that the utilization of UL-256QAM decreased as mobility speed increased. On the passive antenna network, less than 20\% of resource blocks were modulated using 256QAM modulations across all frequency bands due to inadequate channel quality, resulting in an average uplink throughput improvement of only 8.22\%, where most throughput gain is obtained when UE is passing through cell center. However, it was found that enabling UL-256QAM resulted in an increased Block Error Rate (BLER), neglecting some of the throughput gains. On the other hand, on full Massive MIMO deployment, which delivers a superior beamforming performance over conventional passive antennae, sustainable utilization of UL-256QAM can be achieved on the middle-frequency TDD band despite higher path loss with 49.17\% of resource blocks being modulated in 256QAM, pushing spectral efficiency well beyond the limit of UL-64QAM. When considering the TCP latency results, while both TCP congestion control algorithms delivered similar performance in idle scenarios, the latency-based TCP Westwood (TCPW) displayed superior performance when the link was fully loaded, compared to BIC TCP. Enabling UL-256QAM brought an additional improvement in latency across the board. Therefore, for latency-sensitive applications, enabling UL-256QAM and using TCPW on 5G SA will allow TCP applications to enjoy reduced latency.

Since Massive MIMO is only available for middle-frequency TDD bands, it's recommended that MNO with a limited budget first prioritize the roll-out of UL-256QAM on middle-frequency band with Massive MIMO deployment to minimize license fee, then followed by FDD band with passive antennae for lower latency, and throughput gain in the cell center. Finally, if maximum spectral efficiency is desired, then UL-256QAM can be deployed on a middle-frequency TDD band with passive antennae for a slight throughput uplift. As for future work, the effectiveness of UL-256QAM when used with HPUE-enabled UE, which has a higher transmission power budget available, will be evaluated on supported networks to give a full context of uplink performance in the 5G-Advanced era. Especially, with the introduction of the new Power Class 1.5 as a part of 3GPP Release 17 \cite{3GPP_38-306}, which raises the maximum transmission power from 23 dBm to 29 dBm on supported frequency bands such as bands n77 and n78. \looseness=-1

\vspace{-1.5mm}
\section*{Acknowledgement}
\vspace{-0.5mm}

This paper is supported by the commissioned research JPJ012368C03801, National Institute of Information and Communications Technology (NICT), Japan. Additionally, the authors would like to express their gratitude to \textbf{PEI Xiaohong} of Qtrun Technologies for providing Network Signal Guru (NSG) and AirScreen, the cellular network drive test software used for result collection and analysis in this research.

\vspace{-1mm}
\setstretch{0.85}
\renewcommand{\IEEEbibitemsep}{0pt plus 0.5pt}
\makeatletter
\IEEEtriggercmd{\looseness=-1}
\makeatother
\IEEEtriggeratref{1}
\Urlmuskip=0mu plus 1mu\relax

\bibliographystyle{IEEEtran}

\bibliography{bstcontrol,b_reference}

\end{document}